\documentclass[%
 reprint,
 amsmath,amssymb,
 aps,
 prl,
]{revtex4-1}

\usepackage{graphicx}
\usepackage{epstopdf}
\usepackage{dcolumn}
\usepackage{bm}
\usepackage{xcolor}
\usepackage[colorinlistoftodos]{todonotes}
\usepackage{dcolumn}

\begin{document}

\preprint{APS/123-QED}

\title{A New Injection and Acceleration Scheme of Positrons in the Laser-Plasma Bubble Regime}

\author{Z. Y. Xu$^{1}$}
\author{C. F. Xiao$^{1}$}
\author{H. Y. Lu$^{1,2,*}$}
\author{R. H. Hu$^{1 \S}$}
\author{J. Q. Yu$^{1}$}
\author{Z. Gong$^{1}$ }
\author{Y. R. Shou$^{1}$}
\author{J. X. Liu$^{1}$}
\author{C. Z. Xie$^{1}$}
\author{S. Y. Chen$^{1}$}
\author{H. G. Lu$^{1}$}
\author{T. Q. Xu$^{1}$}
\author{R. X. Li$^{3}$}
\author{N. Hafz$^{4}$}
\author{Z. Najmudin$^{5}$}
\author{P. P. Rajeev$^{6}$}
\author{D. Neely$^{6}$}
\author{and X. Q. Yan$^{1,2}$}

\renewcommand{\andname}{\ignorespaces}

\affiliation{$^1$ State Key Laboratory of Nuclear Physics and Technology, and Key Laboratory of HEDP of the Ministry of Education, CAPT, Peking University, Beijing, 100871, China}
\affiliation{$^2$ Collaborative Innovation Center of Extreme Optics, Shanxi University, Taiyuan, Shanxi 030006, China}
\affiliation{$^3$ State Key Laboratory of High Field Laser Physics, Shanghai Institute of Optics and Fine Mechanics, Chinese Academy of Sciences, Shanghai 201800, China}
\affiliation{$^4$ ELI-ALPS, Particle and THz Division, H 6728 Szeged, Hungry, EU}
\affiliation{$^5$ Blackett Laboratory, Imperial College, London SW7 2AZ, UK}
\affiliation{$^6$ Central Laser Facility, Rutherford Appleton Laboratory, Didcot, OX11 0QX, UK}
\affiliation{$*$ Corresponding author: hylu@pku.edu.cn}
\affiliation{$^{\S}$ Current address: School of Physics and Technology, Sichuan University, ronghaohu@scu.edu.cn}


\begin{abstract}

A novel approach for positron injection and acceleration in laser driven plasma wakefield is proposed. A theoretical model is developed and confirmed through PIC simulation. One ring-shaped beam and one co-axially propagating Gaussian beam drive wakefields in a preformed plasma volume filled with both electrons and positrons. The laser's ponderomotive force as well as the charge separation force in the front bucket of the first bubble are utilized to provide  the transverse momenta of injected positrons and those positrons can be trapped by the focusing field and then accelerated by the wakefield. The simulation shows that a relatively high-charge, quasi-monoenergetic positron beams can be obtained.

\end{abstract}
\maketitle

Intense relativistic positron beams are crucial for pair-production in the field of fundamental physics\cite{R1} and violent high-energy astrophysical phenomena\cite{R2}. A stable method to generate intense, mono-energetic and fully tunable positron beams enables experimental study of gamma-ray bursts and black holes\cite{R3, R4}. Bremsstrahlung-based high-energy positrons are usually produced in linear accelerators (LINACs) and synchrotron facilities via propagating the relativistic electron beams through thick, high-Z targets. However, positron beams generated through this method have broadband energy spectra and large transverse divergence, which restrict their applications.

Since the concept of laser wakefield accelerator(LWFA) was first proposed in 1979\cite{R5}, many works have been done on electron acceleration using ultrafast laser systems\cite{R6,R7,R8,R9,R10,R11,R12,R13,R14,R15,R16,R17,R18,R19}. Great progresses have been made in producing high quality electron beams\cite{R6,R7,R8,R10,R11,R12} as well as in boosting the electron energy\cite{R12,R13,R14,R19,R20,R21,R22}. Recently, the generation of sub-hundred MeV positrons has been experimentally demonstrated by hitting these electron beams on high-Z solid targets\cite{R23}. However, the resulting positron beams are limited both in yield and energy. Hybrid schemes have also been proposed and conducted to generate low energy ($E < 20$ MeV) and broad divergence ($\sim$1 rad) positron jets with high positron yields (up to $10^{11}$ per shot)\cite{R24,R25,R26}. High energy and small divergence angle positrons can principally be achieved through laser plasma acceleration of positrons\cite{R27,R28,R29,R30,R31}. However, a scheme for positron injection into these accelerators has yet been absent.

Ultra-short and intense ring-shaped laser propagating in plasma can excite donut-shaped bubbles, which are thought to be suitable structures for positron acceleration\cite{R27,R28,R29,R30,R31}. One typical ring-shaped laser is Laguerre-Gaussian pulse whose azimuthal index is 1 and radial index is 0. The difficulty of positron acceleration in the “bubble” regime\cite{R9} is that positrons, due to their positive charge, are easily expelled away from the bubble in the transverse direction\cite{R32,R33}.

In this Letter, a novel proposal is presented for positrons injection and acceleration, which we tentatively named ``ballistic injection''. A ring-shaped laser beam and a co-axial propagating Gaussian laser beam are employed to create donut and center bubbles in plasma, respectively. In the scheme, the positrons in a bubble can experience the process of scattering, trapping and acceleration before they are emitted or decelerated. In the beginning, the positrons are repelled by the electromagnetic force from the laser tail\cite{R32,R33} like ions scattered by nuclei. The positrons will be injected into the bubbles if the transverse momentum is high enough to let the positrons penetrate the bubble and stay inside. The transverse motion of the positrons is restricted by the focusing electromagnetic field inside the bubble. The longitudinal electric field can accelerate these positrons to relativistic velocities if they stay long enough inside the bubble. 

In this model, two laser beams that have the same temporal profile but with a time delay $\tau$ are employed: $a^{G}(t,r)=a_0^G\exp\left(-\frac{r^2}{r_G^2}-\frac{(t-\tau)^2}{t_0^2}\right)$ for the Gaussian beam and  $a^{R}(t,r)=a_0^R\exp\left(-\frac{(r-r_0)^2}{r_R^2}-\frac{t^2}{t_0^2}\right)$ for the ring-shaped beam, where $r$ is the distance to laser propagation axis and $t$ is time. A positive $\tau$ means the ring-shaped beam is ahead of the center Gaussian beam. For easy understanding, time delay $\tau$ is normalized to $\lambda_p/c$, where $\lambda_p=\frac{2\pi v_g}{\omega_p}$ is the plasma wavelength. Here, $c$ is the speed of light in vacuum, $\omega_p=\sqrt{\frac{4\pi n_e e^2}{m_e}}$ is the plasma frequency, $m_e$ is the electron or positron mass, $e$ is the unit charge of electrons (negative) or positrons (positive), $n_e$ is the electron density, $v_g=c\sqrt{1-n_e/n_c}$ is the group velocity of light in plasma, $n_c$ is the critical density.

The spatial profile of the ring-shaped beam can be characterized by inner radius $r_{in}$ and outer radius $r_{out}$, or ring radius $r_0=(r_{out}+r_{in})/2$ and ring width $r_d=(r_{out}-r_{in})/2$ . The ring-shaped beam is assumed to has the same normalized peak strength $a_0^L$ at $r=r_0$. It can excite a series of donut bubbles with a width of $r_d=2\sqrt{a_0^R}c/\omega_p$ when the matching condition is met, as predicted by nonlinear theory for bubbles\cite{R9}. To clarify the dynamic processes for the scattering and injection, it is better to let the two laser beams interact with their surrounding plasma independently. Thus, the relation $2(\sqrt{a_0^G}+\sqrt{a_0^R})c/\omega_p<r_0$ should be satisfied. In this case, the excited plasma wakefields by Gaussian and ring-shaped laser beams can be analyzed independently.

In this scheme, the plasma region is initially uniformly filled with electrons (density $n_e$), and externally provided positrons (density $n_p$) for simplicity. The positron density $n_p$, is assumed to be much lower than $n_e$, e.g. $n_p=0.01n_e$, to guarantee there are no apparent affection to the plasma field. The contribution of positrons can thus be neglected when analyzing the shape and fields of the plasma bubbles. 

It is key to realize the transfer of positrons between the center and the donut bubbles. The scattering process provides the positrons with transverse momenta, which makes them possible to move from the center bubble to the donut bubble or from the donut bubble to the center bubble for trapping. Otherwise, the positrons will be pushed away by the first half bucket of bubble which is usually overlapped with the rear of the laser pulse. 

For laser propagating along the $x$-axis, under the quasi-static approximation, assume all variables depend on $\xi=x-v_g t$ instead of $x$ and $t$. In the moving frame of the bubble, the currents, densities, and bubble boundaries are then time independent.

The scattering process of positrons is the result of both laser ponderomotive force and charge separation force of the bubbles. The laser ponderomotive force takes the form of $F_p=-\frac{e^2}{4m_e \omega_L^2}\nabla(E^2)=-\frac{m_e c^2}{4}\nabla(a^2)$. The average laser ponderomotive force can be estimated to be $\overline{F_p}=\frac{m_ec^2a_0^2}{4w_0}$, where $w_0$ is the laser spot size.

The total charge separation force on a positron inside the bubble\cite{R34} can be written as $F_c=\frac{m_e\omega_p^2}{2}r$. The total charge separation force on a positron acts as a conservative repulsive force, pointing from the center of bubble to the position of the positron, with a strength proportional to the distance between them. The average charge separation force is estimated to be $\overline{F_c}=\frac{m_e\omega_p^2}{8}\lambda_p$. Thus, the ratio of laser ponderomotive force over charge separation force is determined to be $\delta=\frac{\overline{F_p}}{\overline{F_c}}=\frac{2c^2a_0^2}{\omega_p^2\lambda_pw_0}$.

Given an example case of $a_0 \sim 1$, $n_e \sim 10^{18}\mathrm{/cm^3}$, $w_0 \sim 10\mu$m, thus $\delta<0.1$. It means the charge separation force is the primary scattering factor even in the first bubble, where the laser ponderomotive force also contributes to the scattering effect. The scattering force in the second and the following bubbles comes solely from the charge separation force as the laser is absent due to its short pulse duration. 

The form of the repulsive force keeps the trajectory of a positron in the same plane. $F_c$ can be solved as the Binet equation, giving the trajectory of positrons inside the bubble, which is a hyperbola. For a positron that enters the bubble at coordinate $(\xi_0, y_0)$ and has velocity $v_0$ along the $\xi$-axis. The solutions can be written as:
\begin{equation}
    \begin{cases}
        x &= \frac{v_0}{\omega}\sinh \omega t + x_0\cosh \omega t \\
        y &= y_0\cosh \omega t \\
        \omega &= \sqrt{\frac{1}{2}}\omega_p
    \end{cases}
\end{equation}

The scattering angle can be determined from the above equations.

In the accelerating stage, consider the spherical bubble excited by the Gaussian laser $a^{G}(t,r)=a_0^G\exp\left(-\frac{r^2}{r_G^2}-\frac{(t-\tau)^2}{t_0^2}\right)$. For paraxial electrons and positrons, the one-dimensional fluid model is applied. The scalar potential $\phi(\xi)$ satisfies the Poisson-like equation $\frac{\partial^2\phi}{\partial \xi^2}=k_p^2\gamma_p^2\left[ v_p \left( 1-\frac{1+(a^{G})^2}{\gamma_p^2(1+\phi)^2} \right)^{-\frac{1}{2}} -1 \right]$, where $\gamma_p=(1-v_p^2/c^2)^{-1/2}$ is the Lorentz factor corresponding to the plasma wave phase velocity, and $k_p=\omega_p/v_p$ is the plasma wave vector. The Hamiltonian can be expressed as $H=\sqrt{1+p_{\parallel}^2+p_{\perp}^2} \pm \phi(\xi)$, where the minus sign is for electrons and plus sign is for positrons, respectively. 

Under the canonical transformation $(x, p_{\parallel}) \rightarrow (\xi, p_{\parallel})$, the Hamiltonian becomes $H=\sqrt{1+p_{\parallel}^2+p_{\perp}^2}\pm \phi(\xi)-v_g p_{\parallel}$. This gives several constants of motion. The first is the conservation of transverse canonical momentum, $p_{\perp} \pm a^{G} = \text{const.}$. For positrons initially at rest and far from a sufficiently short laser pulse, this is effectively $p_{\perp}=-a^G$. 

\begin{figure}
    \centering
    \includegraphics[width=0.5\textwidth]{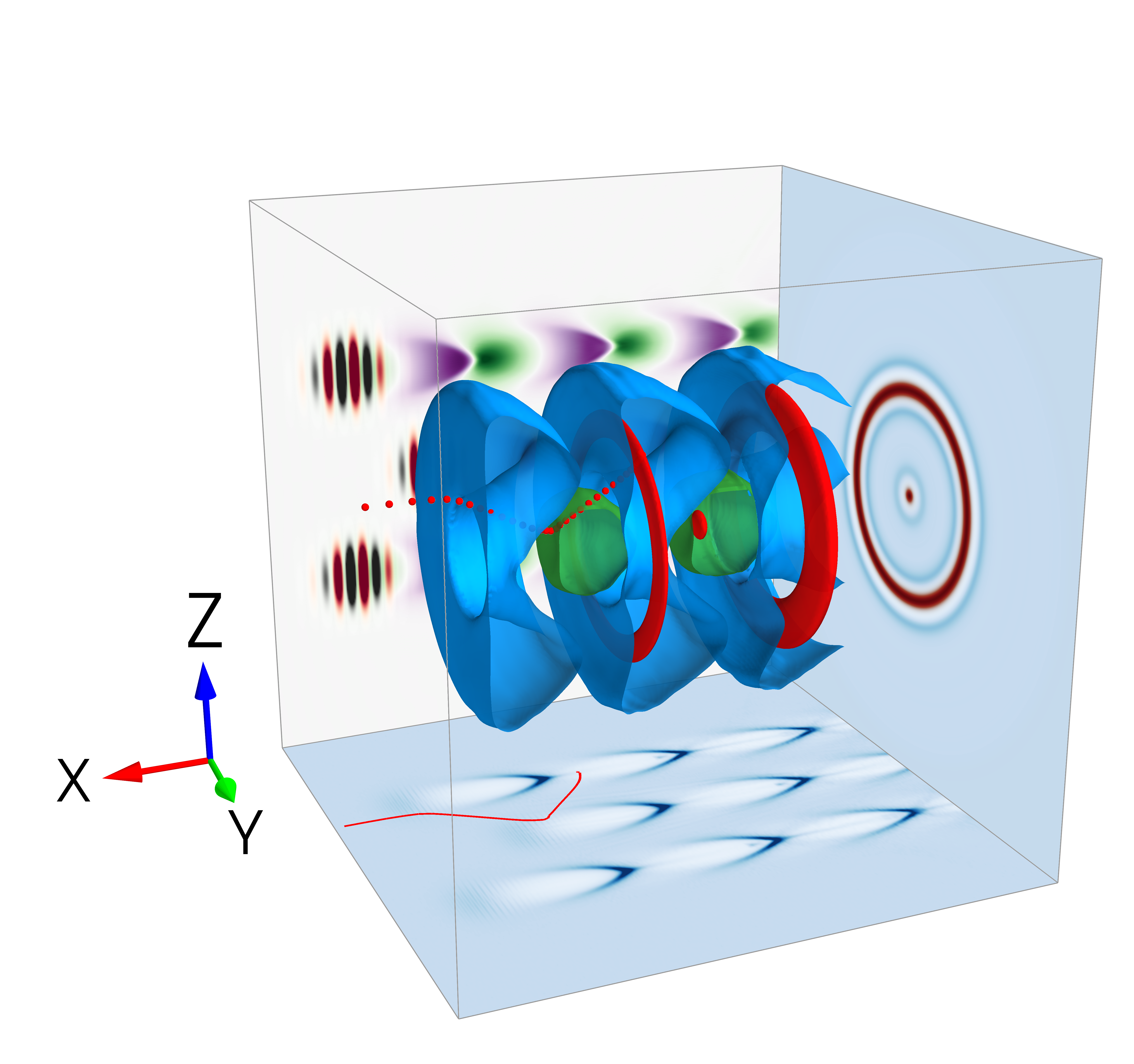}
    \caption{(color online) The concept of the positron ``ballistic injection'' scheme. The blue and green colors are contour surfaces of electron densities of donut and center bubbles, respectively. The red color represents injected positrons. The $x-y$ and $x-z$ planes are transverse slices of the density distribution and the longitudinal electric field $E_x$. The red curve in the $x-y$ plane is the trajectory of an injected positron (corresponding to the projection of red balls in the 3D model). The leading oscillating colors (amber and grey) denotes the laser beams in the $x-z$ plane. The $y-z$ plane is the projection of electron density (blue) and injected positron density (red).}
    \label{FIG:ballistic_injection}
\end{figure}

Another constant of motion is the energy. The solution for longitudinal momentum for a positron with an initial energy is given as $p_{\parallel}=v_p\gamma_p(H_0-\phi) \pm \gamma_p \sqrt{\gamma_p^2(H_0+\phi)^2-p_{\perp}^2-1}$. This solution gives the positron trajectory in $(\xi, p_{\parallel})$ phase space similar to Fig. 1 in Ref. \cite{R32}. The trapped orbits in the second and third bubbles show the potential of actual acceleration of positrons, as the transverse charge separation force acts as a focusing force. However, unlike electron self-injection, positrons initially at rest cannot be self-injected into these areas\cite{R32}. In this injection scheme, the positrons can gain enough transverse momenta, through scattering by other bubbles or the laser pulse, to penetrate the bubble and enter the trapped orbits transversely. The concept is illustrated in Fig. \ref{FIG:ballistic_injection} with a positive time delay as an example. 

The initial injection phase of positrons makes a great difference on the acceleration stage as the dynamics of scattering process is completely different. The delay of the two laser beams can be used as an optimizing tool to ``adjust'' the injection of positrons favoring in the bubbles driven either by the Gaussian beam or by the ring-shaped beam. 

Wide transverse distribution of positrons will lead to the injection of positrons in different bubbles due to different scattering paths, which will result in multi-bunches. Positrons with wide longitudinal distribution, even each of them can experience the same acceleration field in the injected bubble, will lead to the broadening of energy spectra of positrons in each bubble, similar with the case of continuous injection in electron acceleration.

Particle-in-cell (PIC) simulations were conducted using code EPOCH\cite{R35} for detailed understanding the injection and acceleration dynamics. The incident laser propagates along the $x$-axis and is linearly polarized in the $x-y$ plane. Two co-axial propagating laser beams enter the simulation region from the left boundary with $a_0^G=a_0^R=2$. The pulse duration is fixed to 20 fs in full width at half maximum (FWHM) for both beams, and the wavelength is set to 0.8 $\mu$m. The center laser beam has a focused spot size of 10 $\mu$m (FWHM). The ring-shaped beam has a ring radius $r_0=30\mu$m and ring width $r_d=10\mu$m. The ambient plasma electron density is set to $n_e=3\times 10^{18}/\mathrm{cm^3}$, and the first 100 $\mu$m of plasma region is filled with positrons. The positron density is assumed to be $n_p=1\times 10^{16}/\mathrm{cm^3}$ in the simulation, which is already experimentally demonstrated\cite{R24}. The initial temperature of positrons is assumed to be 2 MeV which is considered to be higher enough to prevent annihilation with electrons before they can be accelerated.

The focusing force on positrons in the front bucket of bubbles is the result of high-density electron sheath in the rear of the previous bubbles. Thus, the first donut or center bubble cannot hold positrons as lacking of this focusing force. In the simulations, almost all injected positrons are in the second and third center or donut bubbles, and positrons in the fourth and later bubbles can be neglected.

\begin{figure}
    \centering
    \includegraphics[width=0.5\textwidth]{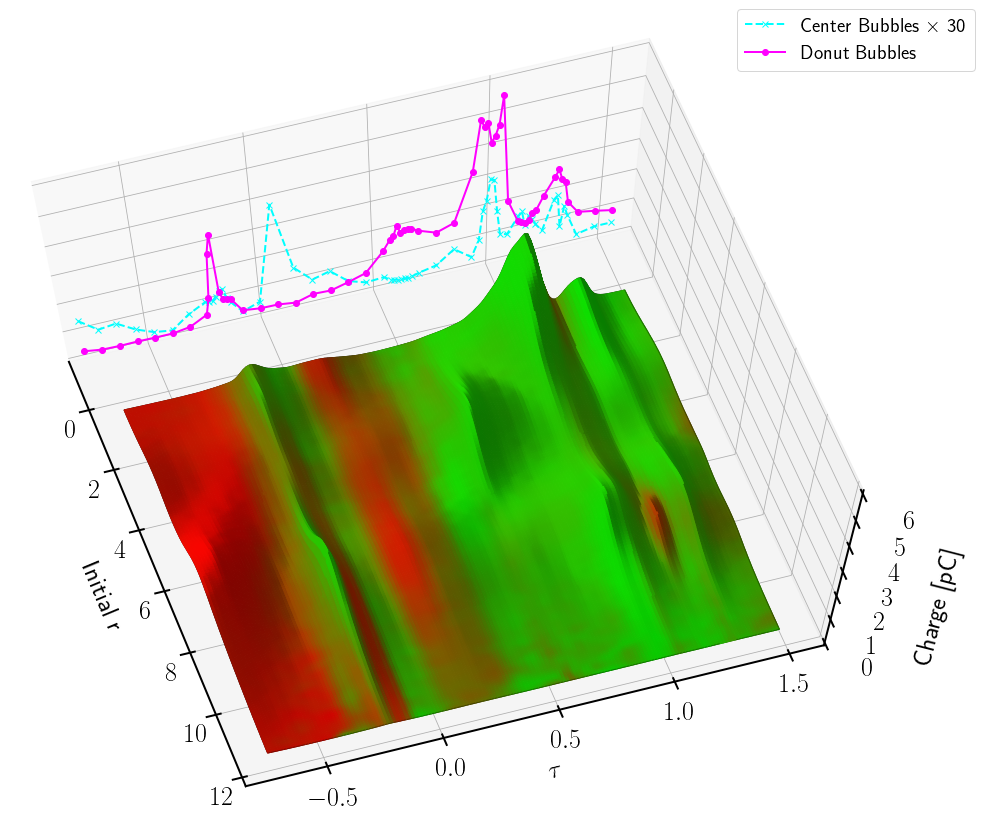}
    \caption{The charge of injected positrons as a function of initial off-axis distance  and delay  in the front three consecutive donut and center bubbles. The value is the total charge, and the color denotes which kind of bubble is dominant, green (red) color means positron charges are mainly from center (donut) bubbles as an example.}
    \label{FIG:injection_count}
\end{figure}

The ponderomotive force of driving laser and the charge separation force in the front bucket of the first bubble behave as defocusing forces on the positrons. The donut bubbles can capture and hold more positrons as their volume are much larger than the center bubbles. In the simulations, the pulse energy of the ring-shaped beam is 8 times higher than that of the Gaussian beam for the same normalized intensity. The volume of the excited donut bubble is roughly 12 times larger than the center bubble considering a 3D configuration. Thus, one donut bubble can hold more positrons than center bubble by considering the 3D charge calculation shown in Fig. \ref{FIG:injection_count}. In the negative delay cases, positrons are captured mainly by the second donut bubble, while the positrons charge increased in the center bubbles in the positive delay cases.  

The initial transverse position $r$ plays a vital role in the injection of positrons as it will determine the scattering trajectory and the following injection position. It is noted that there are two peaks of the charge of injected positrons, with $r$ around 0 and $r_d$(slightly smaller), for several different periodic delays $\tau$. The injection also cuts off beyond $r_d$, because these positrons are scattered outward and leave the bubble region. Thus, the initial distribution of positrons for injection should be confined within the radius $r_d$.

\begin{figure}
    \centering
    \includegraphics[width=0.5\textwidth]{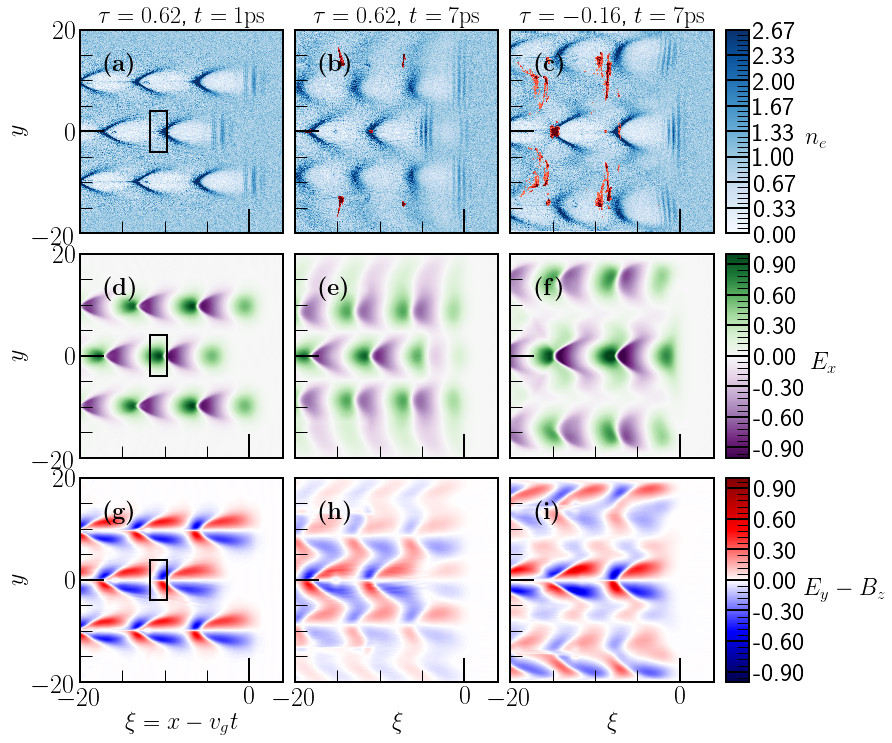}
    \caption{(color online) Simulation results of (a)-(c): electron density distribution $n_e$ (positrons are colored in red), (d)-(f): accelerating field $E_x$ and (g)-(i): focusing field $(E_y-B_z)$ in $t = 1$ ps (near the entry of plasma region) with $\tau = 0.62$ ps ((a), (d), (g)) and $t = 7$ ps (near the exit of plasma region) with $\tau = 0.62$ ps ((b), (e), (h)) and $\tau = - 0.16$ ps ((c), (f), (i)). }
    \label{FIG:sim_result}
\end{figure}

\begin{figure}
    \centering
    \includegraphics[width=0.5\textwidth]{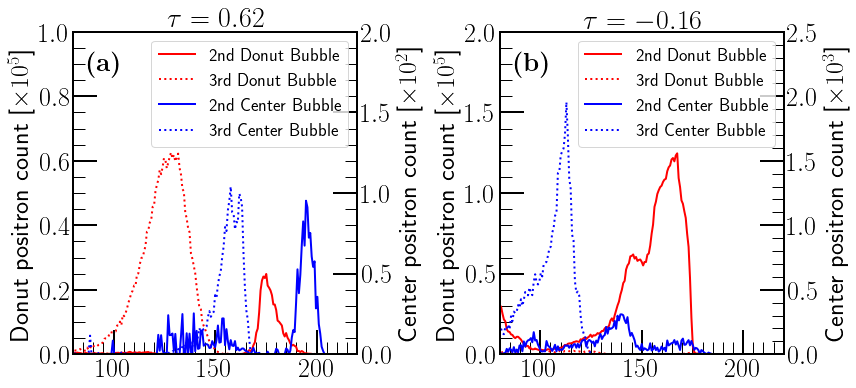}
    \caption{(color online) Typical energy spectra of accelerated positrons in the $2^{\mathrm{nd}}$ and $3^{\mathrm{rd}}$ bubbles of (a) $\tau = 0.62$ and (b) $\tau = -0.16$ at $t = 7$ ps (corresponds to Fig. \ref{FIG:sim_result} (b) and (c), respectively). }
    \label{FIG:spectrum}
\end{figure}

We take two cases with time delays 0.62 and -0.16 for example to describe the injection and acceleration dynamics. The electron density distribution, accelerating and focusing fields at time 1 ps and 7 ps with the two delays are shown in Fig. \ref{FIG:sim_result}. At 1 ps, the positrons in the area, noted with black rectangles, will be focused during acceleration as they experience both positive $E_x$ and negative focusing field gradient. The focusing field can also possibly trap positrons passing through this area, which is the process of injection. The positrons can be accelerated in the bubbles after injection. At 7 ps, most of injected positron are accelerated to high energy (shown in Fig. \ref{FIG:spectrum}), then most of them will start to decelerate. It is worthy to note that 98\% of injected positrons can be accelerated to more than 80 MeV both in the center and donut bubbles in the case of $\tau = 0.62$, while only 37\% can be accelerated to higher than 80 MeV in the delay $\tau = -0.16$, even there are significantly more positrons are injected in the center bubbles.

In the case of $\tau = 0.62$ ps (shown as Fig. \ref{FIG:sim_result} (a)), about 99.8\% of injected positrons (2.06 pC in total) are from the donut bubbles. The relative energy spread ($\delta_E=\Delta E/E_{peak}$, where $\Delta E$ and $E_{peak}$ are the energy range in FWHM and the peak energy of the positron bunch) of the accelerated positrons in the $2^{\mathrm{nd}}$ and $3^{\mathrm{rd}}$ donut bubbles are roughly 6\% and 20\% with peak energies of about 170 MeV and 120 MeV, respectively. The relative energy spread of the positrons are roughly 5\% and 10\% with peak energies of about 200 MeV and 150 MeV in the $2^{\mathrm{nd}}$ and $3^{\mathrm{rd}}$ center bubbles, respectively. Thus, the quality of the accelerated positrons is better in the center bubbles. In the case of $\tau = -0.16$(shown as Fig. \ref{FIG:sim_result} (b)), the accelerated positrons concentrates in the $2^{\mathrm{nd}}$ donut and the $3^{\mathrm{rd}}$ center bubbles due to different scattering patterns, and the positron charge in center bubbles is greatly increased compared with $\tau = 0.62$.

The results suggest optimization of  better relative energy spread or a higher total charge in the acceleration process can be done through time delay $\tau$ control. In the simulation, the magnitude of the accelerating field for positrons is estimated to be the same as electron. Similarly, high energy positrons can also be achieved through increasing the acceleration distance, which needs further optimization of the matching parameters of both laser and plasma densities as those for electrons\cite{R9,R10,R12,R22}.

In this Letter, an injection method for positrons in laser wakefield acceleration is proposed with the requirement of co-axial co-propagating ring-shaped and Gaussian laser pulses. A theoretical model is presented through the description of the dynamical processes experienced by the positrons: scattering, injection and acceleration. The injection method is confirmed through PIC simulations. The simulation shows that a relatively high-charged, quasi-monoenergetic positrons beams (around 200 MeV) can be achieved, the high-energy collimated positron beams are appropriate for applications and further experiments. 

\begin{acknowledgments}
The work has been supported by the National Key Research and Development Program of China (Grant Nos. 2016YFA0401100 and SQ2016zy04003194), the NSFC (Grant No. 11575011), and the Science Challenging Project (Grant No. TZ2017005). Simulations were carried out at High Performance Computing Platform in Peking University.
\end{acknowledgments}


\begin{references}
\bibitem{R1}	C. M. Surko and R. G. Greaves, Phys. Plasmas {\bf 11}, 2333 (2004).
\bibitem{R2}	L. Accardo et al., High statistics measurement of the positron fraction in primary cosmic rays of 0.5-500 GeV with the alpha magnetic spectrometer on the international space station, Phys. Rev. Lett. {\bf 113}, 121101 (2014).
\bibitem{R3}	J. F. C. Wardle, D. C. Homan, R. Ojha and D. H. Roberts, Nature {\bf 395}, 457 (1998).
\bibitem{R4}	G. Weidenspointner, G. Skinner, P. Jean, J. Knodlseder, P. von Ballmoos, G. Bignami, R. Diehl, A. W. Strong, B. Cordier, S. Schanne and C. Winkler, Nature {\bf 451}, 159 (2008).
\bibitem{R5}	T. Tajima and J. M. Dawson, Phys. Rev. Lett. {\bf 43}, 267 (1979).
\bibitem{R6}	J. Faure, Y. Glinec, A. Pukhov, S. Kiselev, S. Gordienko, E. Lefebvre, J.-P. Rousseau, F. Burgy and V. Malka, Nature {\bf 431}, 541 (2004).
\bibitem{R7}	C. G. R. Geddes, C. Toth, J. van Tilborg, E. Esarey, C. B. Schroeder, D. Bruhwiler, C. Nieter, J. Cary and W. P. Leemans, Nature {\bf 431}, 538 (2004).
\bibitem{R8}	S. P. D. Mangles, C. D. Murphy, Z. Najmudin, A. G. R. Thomas, J. L. Collier, A. E. Dangor, E. J. Divall, P. S. Foster, J. G. Gallacher, C. J. Hooker et al., Nature {\bf 431}, 535 (2004).
\bibitem{R9}	W. Lu, C. Huang, M. Zhou, W. B. Mori and T. Katsouleas, Phys. Rev. Lett. {\bf 96}, 165002 (2006).
\bibitem{R10}	J. S. Liu, C. Q. Xia, W. T. Wang, H. Y. Lu, C. Wang, A. H. Deng, W. T. Li, H. Zhang, X. Y. Liang, Y. X. Leng et al., Phys. Rev. Lett. {\bf 107}, 035001 (2011).
\bibitem{R11}	X. M. Wang, R. Zgadzaj, N. Fazel, Z. Y. Li, S. A. Yi, X. Zhang, W. Henderson, Y. Y. Chang, R. Korzekwa, H. E. Tsai et al., Nat. Commun. {\bf 4}, 1988 (2013).
\bibitem{R12}	W. P. Leemans, A. J. Gonsalves, H. S. Mao, K. Nakamura, C. Benedetti, C. B. Schroeder, C. Toth, J. Daniels, D. E. Mittelberger, S. S. Bulanov et al., Phys. Rev. Lett. {\bf 113}, 245002 (2014).
\bibitem{R13}	W. P. Leemans, B. Nagler, A. J. Gonsalves, C. Tóth, K. Nakamura, C. G. R. Geddes, E. Esarey, C. B. Schroeder and S. M. Hooker, Nat. Phys. {\bf 2}, 696 (2006).
\bibitem{R14}	H. Lu, M. Liu, W. Wang, C. Wang, J. Liu, A. Deng, J. Xu, C. Xia, W. Li, H. Zhang et al., Appl. Phys. Lett. {\bf 99}, 091502 (2011).
\bibitem{R15}	J. Faure, C. Rechatin, A. Norlin, A. Lifschitz, Y. Glinec and V. Malka, Nature {\bf 444}, 737 (2006).
\bibitem{R16}	I. Blumenfeld, C. E. Clayton, F. J. Decker, M. J. Hogan, C. K. Huang, R. Ischebeck, R. Iverson, C. Joshi, T. Katsouleas, N. Kirby et al., Nature {\bf 445}, 741 (2007).
\bibitem{R17}	W. Lu, M. Tzoufras, C. Joshi, F. S. Tsung, W. B. Mori, J. Vieira, R. A. Fonseca and L. O. Silva, Phys. Rev. Spec. Top.- Accel. Beams. {\bf 10}, 061301 (2007).
\bibitem{R18}	S. Kneip, S. R. Nagel, S. F. Martins, S. P. D. Mangles, C. Bellei, O. Chekhlov, R. J. Clarke, N. Delerue, E. J. Divall, G. Doucas et al., Phys. Rev. Lett. {\bf 103}, 035002 (2009).
\bibitem{R19}	T. Kameshima, W. Hong, K. Sugiyama, X. Wen, Y. Wu, C. Tang, Q. Zhu, Y. Gu, B. Zhang, H. Peng et al., Appl. Phys. Express {\bf 1}, 066001 (2008).
\bibitem{R20}	C. E. Clayton, J. E. Ralph, F. Albert, R. A. Fonseca, S. H. Glenzer, C. Joshi, W. Lu, K. A. Marsh, S. F. Martins, W. B. Mori et al., Phys. Rev. Lett. {\bf 105}, 105003 (2010).
\bibitem{R21}	S. Steinke, J. van Tilborg, C. Benedetti, C. G. Geddes, C. B. Schroeder, J. Daniels, K. K. Swanson, A. J. Gonsalves, K. Nakamura, N. H. Matlis et al., Nature {\bf 530}, 190 (2016).
\bibitem{R22}	A. J. Gonsalves, K. Nakamura, J. Daniels, C. Benedetti, C. Pieronek, T. C. H. de Raadt, S. Steinke, J. H. Bin, S. S. Bulanov, J. van Tilborg et al., Phys. Rev. Lett. {\bf 122}, 084801 (2019).
\bibitem{R23}	G. Sarri, W. Schumaker, A. Di Piazza, M. Vargas, B. Dromey, M. E. Dieckmann, V. Chvykov, A. Maksimchuk, V. Yanovsky, Z. H. He et al., Phys. Rev. Lett. {\bf 110}, 255002 (2013).
\bibitem{R24}	H. Chen, S. C. Wilks, J. D. Bonlie, E. P. Liang, J. Myatt, D. F. Price, D. D. Meyerhofer and P. Beiersdorfer, Phys. Rev. Lett. {\bf 102}, 105001 (2009).
\bibitem{R25}	H. Chen, S. C. Wilks, D. D. Meyerhofer, J. Bonlie, C. D. Chen, S. N. Chen, C. Courtois, L. Elberson, G. Gregori, W. Kruer et al., Phys. Rev. Lett. {\bf 105}, 015003 (2010).
\bibitem{R26}	H. Chen, F. Fiuza, A. Link, A. Hazi, M. Hill, D. Hoarty, S. James, S. Kerr, D. D. Meyerhofer, J. Myatt et al., Phys. Rev. Lett. {\bf 114}, 215001 (2015).
\bibitem{R27}	A. S. Firouzjaei and B. Shokri, Phys. Plasmas {\bf 24}, 013107 (2017).
\bibitem{R28}	L. L. Yu, C. B. Schroeder, F. Y. Li, C. Benedetti, M. Chen, S. M. Weng, Z. M. Sheng and E. Esarey, Phys. Plasmas {\bf 21}, 120702 (2014).
\bibitem{R29}	J. Vieira and J. T. Mendonca, Phys. Rev. Lett. {\bf 112}, 215001 (2014).
\bibitem{R30}	S. Gessner, E. Adli, J. M. Allen, W. An, C. I. Clarke, C. E. Clayton, S. Corde, J. P. Delahaye, J. Frederico, S. Z. Green et al., Nat. Commun. {\bf 7}, 11785 (2016).
\bibitem{R31}	C. B. Schroeder, C. Benedetti, E. Esarey and W. P. Leemans et al., Phys. Plasmas {\bf 20}, 123115 (2013).
\bibitem{R32}	T. Esirkepov, S. V. Bulanov, M. Yamagiwa and T. Tajima, Phys. Rev. Lett. {\bf 96}, 014803 (2006).
\bibitem{R33}	W. Lu, C. Huang, M. M. Zhou, W. B. Mori and T. Katsouleas, Phys. Plasmas {\bf 12}, 063101 (2005).
\bibitem{R34}	I. Kostyukov, A. Pukhov and S. Kiselev, Phys. Plasmas {\bf 11}, 5256 (2004).
\bibitem{R35}	T. D. Arber, K. Bennett, C. S. Brady, A. Lawrence-Douglas, M. G. Ramsay, N. J. Sircombe, P. Gillies, R. G. Evans, H. Schmitz, A. R. Bell and C. P. Ridgers, Plasma Phys. Control. Fusion {\bf 57}, 113001 (2015).
\end{references}
\end{document}